\documentclass[10pt]{article}
\usepackage{amsmath}
\usepackage{amssymb}
\usepackage{graphicx}

\usepackage[numbers,sort&compress]{natbib}

\usepackage{color}
\usepackage{epsfig}
\usepackage{subfigure}
\usepackage{bm}
\usepackage{latexsym}
\usepackage{listings}
\usepackage{wasysym}
\usepackage{mathtools}

\usepackage[labelfont=bf,labelsep=period,justification=raggedright]{caption}

\makeatletter
\renewcommand{\@biblabel}[1]{\quad#1.}
\makeatother

\date{}

\begin{document}

\begin{flushleft}
{\Large
\textbf{Piezoelectric Drop-on-Demand Inkjet Printing of Rat\vspace{0.08 in}\\
Fibroblast Cells: Survivability Study and Pattern Printing}\vspace{0.2 in}
}
\\
Er Qiang Li$^{1}$,
Eng Khoon Tan$^{2}$,
Sigurdur T. Thoroddsen$^{1,\ast}$\vspace{0.1 in}
\\
{\bf 1} Division of Physical Sciences and Engineering, King Abdullah University of Science and Technology (KAUST), Thuwal, Saudi Arabia
\\
{\bf2} Mechanical Engineering, National University of Singapore, Singapore
\\
$\ast$ E-mail: sigurdur.thoroddsen@kaust.edu.sa
\end{flushleft}

\section*{Abstract}
A novel piezoelectric, drop-on-demand (DOD) inkjet system has been developed and used to print
L929 rat fibroblast cells.  We investigate the survivability of the cells subjected to the
large stresses during the printing process. These stresses are varied by changing the diameter
of the orifice (36 to 119 $\mu$m) through which the cells are dispensed, as well as changing
the electrical pulse used to drive the piezoelectric element.  It is shown that for the
smallest 36 $\mu$m diameter orifice, cell survival rates fall from 95\% to approximately 76\%
when the ejection velocity is increased from 2 to 16 m/s.  This decrease in survival rates
is less significant when the larger orifice diameters of 81 $\mu$m and 119 $\mu$m are used.
Analysis shows that there is a clear inverse relationship between cell survival rates and the
mean shear rates during drop formation.  By using the same printing set-up, fibroblast cells
are printed onto alginate and collagen into patterns.  Printed cells are cultured over a
period of days to verify their long-term viability.  Fibroblasts printed onto the collagen
are found to successfully adhere, spread and proliferate, subsequently forming a denser
patterns after 5 days in culture.  Cell agglomeration is found to affect the printing
performance, especially for the printhead with the smallest orifice, leading to frequent
clogging of the nozzle.  We also study the number of cells in each droplet, when printed
under optimal conditions.  The probability density of this number follows a binomial distribution,
which consistent with a uniform distribution of cells in the medium and within the printhead.

\section{Introduction}

One of the aims of tissue engineering is to position cells into 3-dimensional structures
and arrange them in a specific pattern. The generation of such structures forms the basis
of tissue regeneration and possibly, organ building \cite{Calvert2007}. Inkjet printing is
a suitable candidate for this purpose. It has been used successfully in a similar manner
for automated rapid prototyping technology which precisely positions droplets onto a
substrate \cite{Derby2012,Merrin2007,Cui2009,Cui2012,Boland2006,Ferris2013,Guillotin2011,Xu2013,Cohen2009,Melchels2012}.
Successful organ printing should most importantly fulfill the following three requirements:
cells should be accurately placed into desired patterns; cells should maintain their survivability
after printing and cells should get rapid and continuous deposition and solidification onto a
thermo-reversible gel \cite{Mironov2003}. The gel must ensure sufficient transport of nutrients,
provide adequate mechanical support and allow cell adhesion. The gel should also allow spreading
and proliferation of multiple cell types \cite{Fedorovich2007} and must subsequently degrade
in a regular and predictable fashion \cite{Augst2006}.

To date, many different cell types have been printed successfully by different printing methods
and their viability has been
verified \cite{Chang2008,Xu2004,Xu2005,Chen2006,Barron2005,Nakamura2005,Saunders2008,Xu2006,Xu2013}.
By dispensing human fibroblast cells through a 60 $\mu$m nozzle, a
comprehensive study was carried out to investigate the relationship between cell survivability
and the inkjet printing parameters \cite{Saunders2008}. Their study supported previous
claims \cite{Xu2004,Xu2005,Nakamura2005,Xu2006} that cell
survivability was not significantly affected by the printing process since cell survival
rates only fell from 98\% to 94\% when the excitation pulse voltage was increased from 40 to
80 V. A similar high printed cell viability rate of 89\% was also reported when
Chinese hamster ovary cells were dispensed by a thermal inkjet printer \cite{Cui2010}.
However, in their study, the printing process was all carried out using a modified
commercial printer, thus limiting their experiments to a fixed nozzle diameter (60 $\mu$m \cite{Saunders2008}
or 48 $\mu$m \cite{Cui2010}) and a small range of the drop velocities.
This limitation may be of importance, because the shear stresses, which are expected
to be the main factor in the killing of cells during the printing process,
are proportional to the velocity gradients within the nozzle.

To eliminate the above two limitations, a squeeze mode piezoelectric drop-on-demand (DOD)
printing system was developed in-house for this study. A much larger range of droplet
velocities was obtained, compared to the former studies \cite{Saunders2008,Cui2010}. Furthermore,
the improved design of the printhead allows us to change the glass nozzles while using the
same piezo-electic printhead \cite{Li2009,Li2010}, thus enabling us to investigate the effects of varying the
orifice diameters.

It will be shown that by using a small diameter nozzle and a high excitation voltage,
the printing process generates large enough shear stresses to cause significant decrease
in the cell survival rates.  In fact, shear stresses have been studied extensively to
predict the damage of animal cells suspended in various laminar or turbulent
flows \cite{Chisti2001,Born1992,Zoro2008,Zhang1993}. Herein we
provides a quantitative study of these effects on cell survivability in piezoelectric DOD inkjet printing.

The number of cells in each printed droplet will be one factor in optimizing cell printing,
as empty droplets may be undesirable \cite{Yamaguchi2012}. We have studied the probability distribution of cell
numbers to ascertain desirable mean cell concentration in the medium, to avoid cell-less
droplets.

Herein we first printed L929 rat fibroblast cells onto alginate to form patterns. Alginate
has been increasingly utilized in tissue engineering to support encapsulated cells and to
regulate cells function, in a manner similar to the extracellular matrices of mammalian
tissues \cite{Kong2003}. Alginate¡¯s popularity comes from its advantages of
biocompatibility, nonimmunogenicity \cite{Shapiro1997} and gentle gelling
behavior \cite{Klock1997}. However, the major limitation to its use as extracellular
matrices is that alginate does not mediate mammalian cell adhesion \cite{Augst2006,Alsberg2001}.
To promote cell adhesion within alginate gel, ligands such as
arginine-glycine-aspartic acid (RGD) \cite{Fedorovich2007,Mann1999,Massia1990,Olbrich1996},
GRGDY \cite{Rowley1999}, KGD and VAPG \cite{Mann1999} have been used.
Collagen is another widely used hydrogel with a
number of advantages including biodegradability, low immunogenicity and controllable
stability. Furthermore, collagen contains cell adhesion domain sequences such as RGD,
which facilitate cell adhesion for anchorage-dependent cell types \cite{Patino2002,Silver1992}.
Therefore, L929 cells were later delivered onto collagen to print patterns.

\section{Results}

\begin{figure}[!ht]
\begin{center}
\includegraphics[width=3.27in]{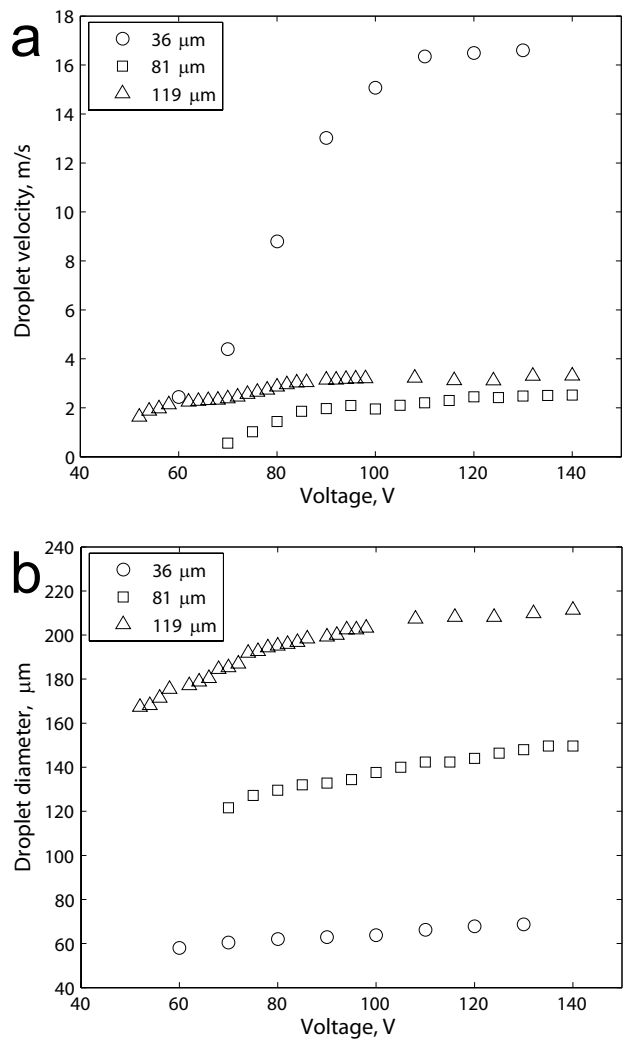}
\end{center}
\caption{
{\bf Influence of excitation pulse amplitude on printed (a) droplet velocity and (b) droplet diameter.}
The orifice diameters of the nozzles used were 36, 81 and 119 $\mu$m.
}
\label{Fig1}
\end{figure}

\subsection{Cell Printing}

L929 cells were printed into petri dishes containing the live-dead assay solution, through
3 different orifices with the diameters of 119 $\mu$m, 81 $\mu$m and 36 $\mu$m. Printing
was carried out over a range of excitation pulse amplitudes from 52 to 140 V, while the
rising/falling time was kept constant at 3 $\mu$s and the dwell time, i.e. the time
duration of the excitation pulse, was kept at 70 $\mu$s. The driving frequency was held
constant at 1.5 kHz. Each sample was printed for approximately 20 s. The concentration
of the cell suspension was about 1 million cells per ml.

The initial average number of cells inside each droplet is fairly independent of the voltage
used to drive the piezo-element. This was verified by observations under the microscope done
within 2 hours of the printing, i.e. before proliferation occurs. This result is consistent
with the existing study \cite{Saunders2008}; however, the average number of cells depends
strongly on the nozzle/drop size, as discussed below.

Figure~\ref{Fig1} shows the effect of the excitation pulse (which is imparted to the piezoelectric
actuator) on the droplet velocity and droplet diameter. It is shown that for all of the three
nozzle sizes, drop velocity and droplet diameter increase with the increase of excitation pulse.
This increase in droplet velocity is especially pronounced for the 36 $\mu$m nozzle, where
droplet velocity increases from 2.4 to 16.6 m/s as the driving voltage increases from 60 V
to 130 V. However, for the 36 $\mu$m nozzle, small satellite droplet is generated once the
driving voltage exceeds 70 V. To avoid satellite formation typical DOD inkjet printing cannot
generate such high velocities as used herein. In fact, for a specific nozzle size and pulse
duration, there exists a critical pulse amplitude, above which satellite droplets will be
produced \cite{Dong2006,Li2010}. When a satellite droplet is
generated, the drop velocity is determined based on the main droplet. The presence of the
small satellite droplets is of no relevance to the survival study, but may interfere with
pattern printing.

\subsection{Cell Survivability: Effects of the Mean Shear Rate}

\begin{figure}[!ht]
\begin{center}
\includegraphics[width=3.27in]{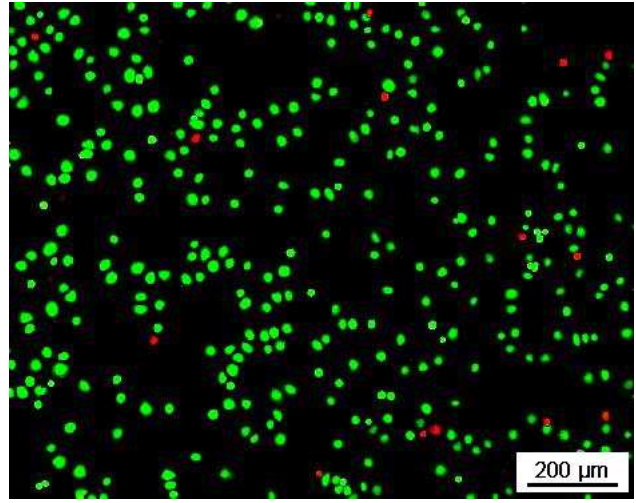}
\end{center}
\caption{
{\bf Graph showing a 96\% survival rate of L929 rat fibroblast cells stained with Calcein AM and Ethidium homodimer (EthD-1).}
Printed with an excitation pulse amplitude of 116 V, at a frequency of 1.5 kHz,
with rising and falling times of 3 $\mu$s. The orifice used was 119 $\mu$m.
}
\label{Fig2}
\end{figure}

Cell survivability after printing was quantitatively investigated by using the LIVE-DEAD
Viability/Cytotoxicity Kit. Figure~\ref{Fig2} shows a typical stained sample which has a
96\% survival rate.

\begin{figure}[!ht]
\begin{center}
\includegraphics[width=3.27in]{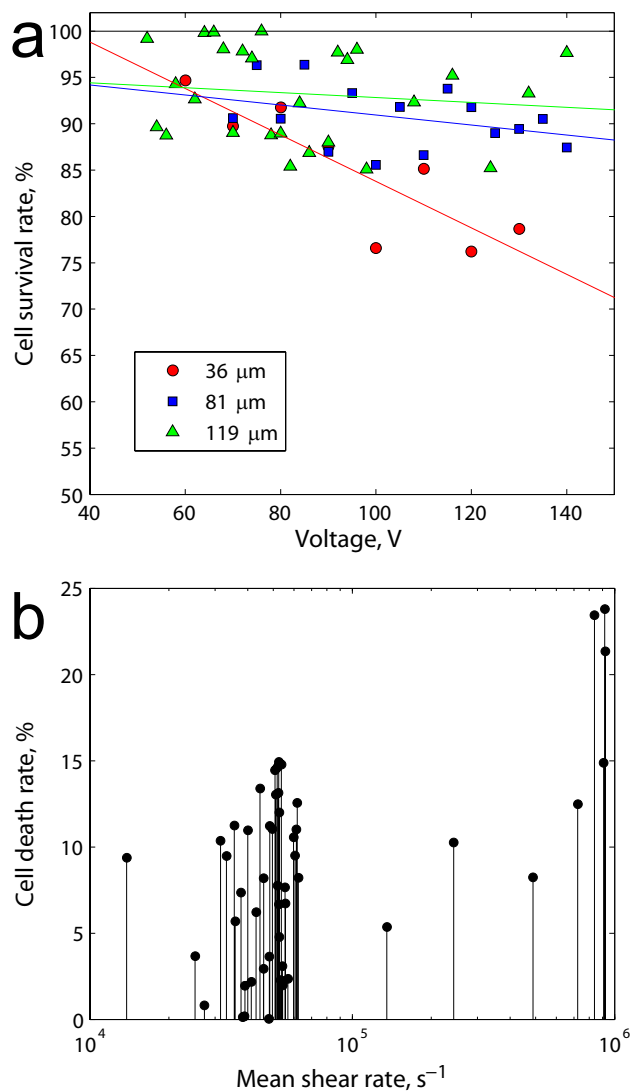}
\end{center}
\caption{
{\bf Survivability of printed cells.}
(a) Mean cell survival rate against excitation pulse amplitude. Samples printed through orifices
with the diameter of 36, 81 and 119 $\mu$m, with excitation pulse amplitude from 52 to 140 V, at
frequency of 1.5 kHz, with rising and falling times of 3 $\mu$s. Each cell survival rate data was
the average value from 5 replicates.
(b) Percentage of cell death against the mean shear rate.
Samples printed through orifices with the diameter of 36, 81 and 119 $\mu$m. Each cell death rate
data was the average value from 5 replicates.
}
\label{Fig3}
\end{figure}

Figure~\ref{Fig3}a shows the mean cell survival rates against excitation pulse amplitude for three
different orifices, to compare the effects of orifice sizes on cell survivability. It is
shown that for the 36 $\mu$m nozzle, the survival rate falls from 95\% to 78\% as the
excitation pulse increased from 60 to 130 V. The lowest survival rate of 76\% is observed
when the highest voltage is approached. The excitation pulse amplitude represents the power
for the piezoelectric actuator to dispense the droplets and this power directly affects the
droplet velocity and thereby the shear stress in the liquid. It is also shown that survival
rates fall from 99\% to 85\% for the 119 $\mu$m orifice and from 96\% to 85\% for the
81 $\mu$m orifice. It can be seen that for the bigger orifices, especially the 119 $\mu$m
one, the printing did not produce a significant reduction in cell survivability as the
excitation pulse amplitude is increased. This may be due to the fact that the cells used
here were much smaller than the two bigger orifices. The round-shaped L929 rat fibroblast
cells are measured to have a diameter of approximately 20 $\mu$m. It is known that shear
stress is proportioned to the velocity gradient in radial direction, thus the highest
shear stress is always generated near the wall region during droplet dispensing. For the
larger nozzles the likelihood of the cell moving next to the wall is reduced, on average
the cells will therefore experience less shear stresses, which would ultimately lead to
a higher survival rate.

Comparing the 3 different trendlines in Figure~\ref{Fig3}a we conclude that it is not the strength
of the electric field which directly affects cell survival rate, rather it is the fluid
shear stress. Due to the highly transient nature of the flow driven through the nozzle,
we lack detailed knowledge of the velocity profile within the droplet at
the nozzle tip. Here we use the mean shear rate, which can be estimated by
$\dot{\varepsilon} = v/r$, where $v$ and $r$ are the drop velocity and the nozzle radius, respectively,
as a substitute for effective shear stresses. Figure~\ref{Fig3}b shows the percentage of cells
that died against the mean shear rate during the printing. It is shown that the cell death
rate increases approximately from 5\% to 24\% as the mean shear rate increases from
$1.4 \times 10^4$ $s^{-1}$ to $9.2 \times 10^5$ $s^{-1}$.  The trend is more evident for
the last eight data points which correspond to the results for the 36 $\mu$m orifice.
The results clearly show that cell death does occur during the printing, especially
under the effects of high shear rates that are above $5 \times 10^5$ $s^{-1}$.

One can roughly estimate the displacement thickness of the boundary layer $\delta$ using the time
duration of the piezo-signal $T$ and the kinematic viscosity of the liquid $\nu$, with the
well-known approximation \cite{Batchelor1967}, $\delta = 1.72 \left( \nu T \right) ^{1/2}$.
The total duration of the signal is $T$ = 76 $\mu$s and the viscosity of the cell ink is
very similar to that of water, i.e. $\nu = 10^{-3}$ $m^{2}/s$, which gives $\delta = 15$ $\mu$m.
The two boundary layers therefore span 30 $\mu$m, which is close to the diameter of the
smallest nozzle. The large velocity gradients inside the boundary layers of this nozzle are therefore
likely to submit many of the cells to the high shear stresses. The geometry of the
converging nozzle will certainly affect the true thickness of these boundary layers,
but this simple calculation suggests that their size becomes quite significant for the
smallest nozzle diameter of 36 $\mu$m.

\begin{figure}[!ht]
\begin{center}
\includegraphics[width=3.8in]{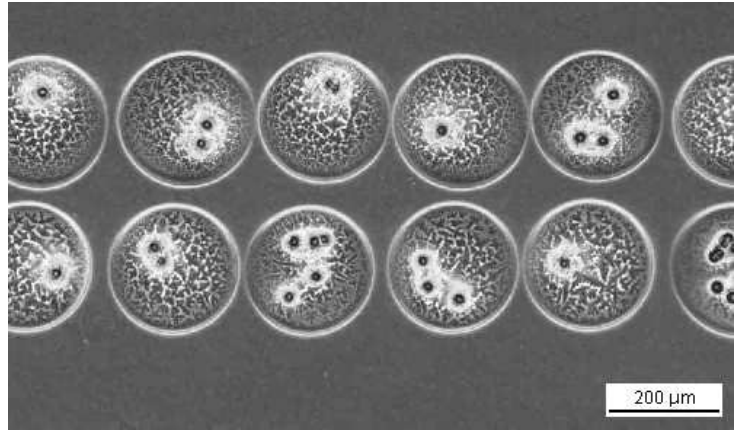}
\end{center}
\caption{
{\bf Droplets printed onto a dry substrate from a suspension with a concentration of 2 $\bf \times$ $\bf 10^6$ cells per ml.}
Each droplet contains 1 to 5 cells. The orifice diameter of the nozzle used was 60 $\mu$m.
}
\label{Fig4}
\end{figure}

\subsection{The Number of Cells in Each Droplet}

Having investigated cell viability from the inkjet printing, the next step in optimizing
the use of such printing in tissue engineering is to uniformly position cells in desired
configurations.

Figure~\ref{Fig4} shows two adjacent straight lines printed with a 60 $\mu$m nozzle. The space
between the lines is around 30 $\mu$m. There are between 1 to 5 cells observed in each
droplet. This large deviation in cell numbers highlights the random distribution of the
cells inside the medium when it reaches the nozzle, from which the droplets are dispensed.
Therefore, a large average cell concentration will be needed in the suspension to guarantee
at least one cell per droplet, with a certain high probability. To investigate the
associated probabilities we performed a set of separate experiments described in what follows.

The number of cells in each droplet can be thought of as a random variable, whose
distribution can then be estimated using basic probability theory.
We assume that the distribution of cells in the original medium is random with a uniform
probability density. With this assumption the printing simply represents random sampling
of the liquid volume in the reservoir, with a sphere of the same volume as the droplet $\Omega_d$.
The aim is to determine the probability that a certain number of cells are present in this
volume. We can formulate this in terms of a Bernoulli sequence of trials \cite{Papoulis1965}.
Each trial consists of randomly assigning the position of the center of one cell inside
the whole liquid volume of the media $\Omega_m$. Successful trial occurs when the cell
lands inside our specified droplet. Then the probability of getting $k$ cells into a
specific droplet in $n$ trials becomes

\begin{equation}
P_n \left( k \right) = e^{-np} \frac{ \left( np \right)^k }{k!}
\label{Pn}
\end{equation}
where $p$ is the probability of success in each trial, $N$ is the average cell
concentration per unit volume, and the product $np = N \Omega_d$ now corresponds to the average
number of cells per droplet volume, which we denote by $N_d$.

\begin{figure}[!ht]
\begin{center}
\includegraphics[width=3.27in]{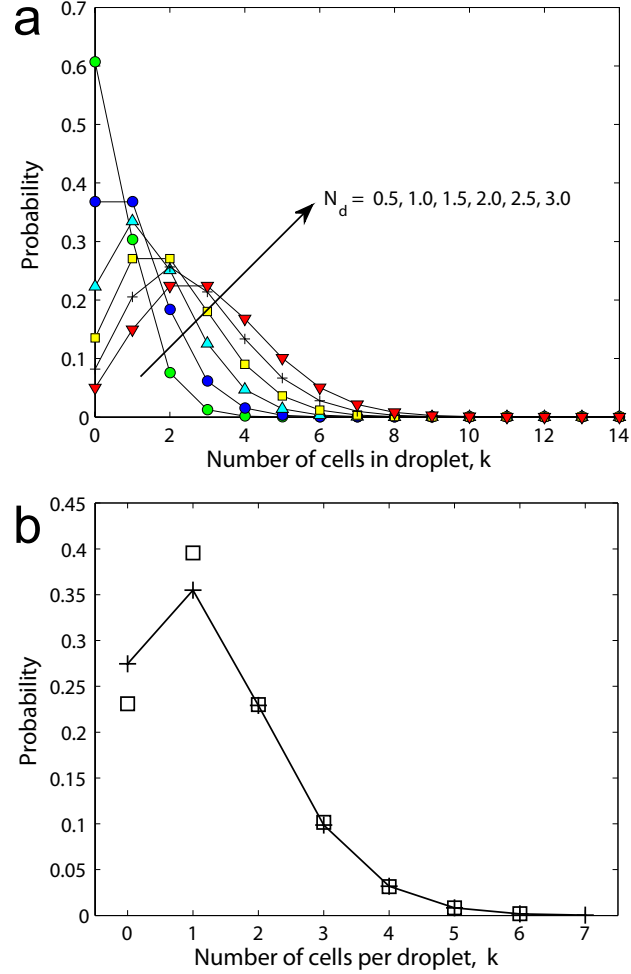}
\end{center}
\caption{
{\bf Graph showing the probability density distribution of the number of cells in each droplet.}
(a) For a range of different average cell concentration in the cell medium, from $N_d$ = 0.5,
1.0, 1.5 $\cdots$ 3.0 cells per droplet.
(b) Experimentally determined cell number distribution.  The ($\square$) stands for the experiment results and (+) stands
for the values calculated from Eqn.(\ref{Pn}). Determined from microscope counting of cells in 800 droplets
dispensed within the first 4 minutes. $N_d$ is 1.29.
}
\label{Fig5}
\end{figure}

Figure~\ref{Fig5}a shows this probability density function for a few different values of $N_d$,
highlighting the variability of the number of cells in different droplets. As the
concentration increases the most likely number of cells shifts to larger $k$, while the
distribution also widens. This result is consistent with previous studies \cite{Merrin2007,Moon2011}.
The figure shows clearly, that there is a finite probability
of producing droplets containing no cells. The above Eqn.(\ref{Pn}) shows that the
probability of empty droplets is

\begin{equation}
P_n \left( k = 0 \right) = e^{-np}
\label{Pnk0}
\end{equation}
which reduces exponentially with higher cell concentration in the medium. Using this
formula, we can see that to guarantee, with a 99\% probability, that each droplet contains
at least one cell the average cell density in the solution must be above $N_d$ = 4.6 cells/drop.

\begin{figure}[!ht]
\begin{center}
\includegraphics[width=4.0in]{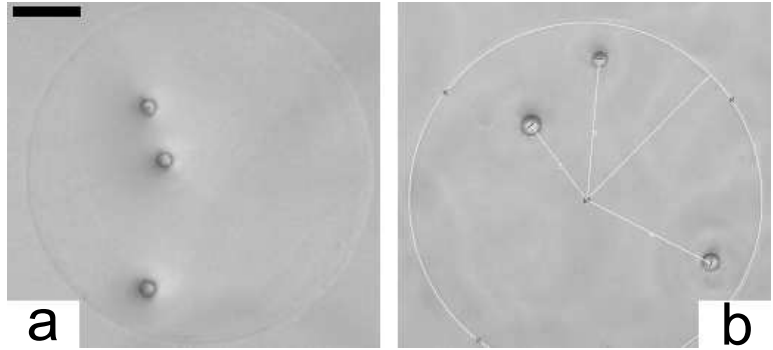}
\end{center}
\caption{
{\bf Images of printed cells.}
(a) Cells inside dried droplet residues. The scale bar is 50 $\mu$m.
(b) Schematic showing the measurement of the radial location of each cell, away from the center
of the dried droplet residue.
}
\label{Fig6}
\end{figure}

To investigate the validity of our assumptions we carried out a set of experiments where
thousands of droplets were printed onto dry petri dishes. The diameter of the nozzle used
was 130 $\mu$m, giving droplet diameter of 170 $\mu$m. The resulting diameter of the dry
droplet residue was 268 $\mu$m. Therefore the average height of the liquid film was 46 $\mu$m.
The drops dried out within about 1 min, leaving the cells encased inside the remaining residue
of dried medium, as is shown in Figure~\ref{Fig6}a. The number of cells inside each drop was then
counted under the microscope. Figure~\ref{Fig5}b compares the distribution of the number of cells
inside each droplet with the theory in Eqn.(\ref{Pn}). The theory shows excellent agreement except
that we observe slightly fewer empty drops that predicted and slightly more droplets
containing only one cell. The theory shows perfect agreement for $k \geqslant 2$. Keep
in mind that there are no free parameters in this relationship, with the mean cell
concentration $N_d$ = 1.29 coming directly from the experimental results. This excellent
agreement with the theory is expected to become even better when the average number of
cells in the medium increases.

\subsection{The Location of Cells inside Droplets}

\begin{figure}[!ht]
\begin{center}
\includegraphics[width=3.27in]{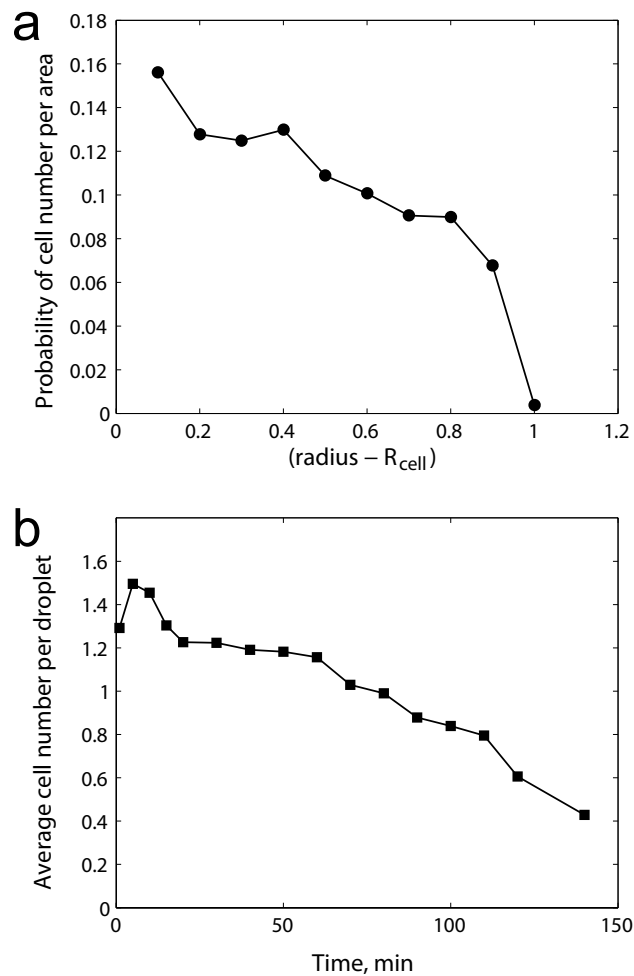}
\end{center}
\caption{
{\bf Two parameters which affect the printing of cell patterns.}
(a) The probability of cell location within the dried droplet splatter.
(b) The average number of cells per droplet vs. time from start of printing. Printing was
carried out continuously over a period of 2.5 hours, at a 120 Hz driving frequency.
}
\label{Fig7}
\end{figure}

The spatial distribution of the cells within the dried drop was also studied. Figure~\ref{Fig6}b
shows how we measured the radial location of each cell, away from the center of the dried
droplet residue. We first verified that the horizontal motion of the substrate, during
the printing, does not move the cells towards a specific direction. This might be
introduced by the effective angle of impact of the droplet, which is always less
than 6 $^{\circ}$ from the vertical. The evaporation of the liquid could also introduce
capillary-driven motions of the cells to the edge of the drop, as is well-known from the
everyday experience of coffee stains \cite{Deegan1997}. This was not observed in the
resulting distribution of cells in the spot, which is normalized by the area. Figure~\ref{Fig7}a
shows that the cells are most likely to be located near the center, with clear reduction
in cell numbers near the edge. This might be explained if the thin lamella of liquid which
is generated by the impact and precedes the spreading, is of similar thickness as the cells.
This is likely to occur in our setup, as the Reynolds number of the
impacts $Re = \rho D U_i / \mu \approx 170$, suggesting a weak lamella traveling
along the substrate \cite{Thoroddsen1998}. Here $\rho$ and $\mu$ are respectively the liquid density and dynamic
viscosity. $U_i$ and $D$ are the impact velocity and droplet diameter.

Figure~\ref{Fig7}b shows the long-time evolution of the average number of cells in each droplet.
The cell number is fairly uniform for the first hour and then reduces at approximately a
uniform rate, which is probably due to slow coagulating or settling of the cells in the
liquid reservior.

\subsection{Printing Patterns}

\begin{figure}[!ht]
\begin{center}
\includegraphics[width=3.6in]{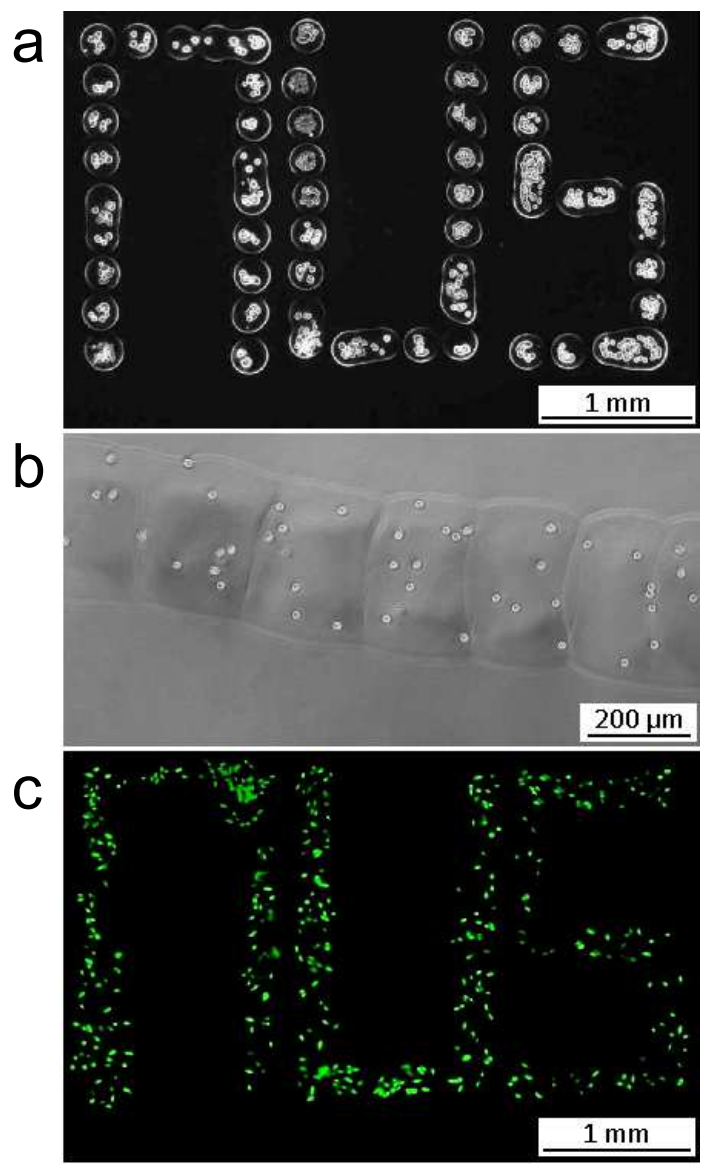}
\end{center}
\caption{
{\bf Image showing results of cell pattern printing.}
(a) Cells printed onto a dry Petri-dish, forming an ``NUS'' pattern. Each droplet contains
2 to 6 cells.
(b) A continuous line of overlapping drops with around 6 to 8 cells/drop in the crosslinked gel.
(c) Cells printed onto a collagen gel, forming an ``NUS'' pattern. Picture taken 5 day after printing.
The orifice diameter of the nozzle used was 60 $\mu$m.
}
\label{Fig8}
\end{figure}

For the printing of cell patterns, a primitive manual micro-meter x-y-stage is used,
for a proof-of-concept demonstration. Figure~\ref{Fig8}a shows cells fired onto a dry petri-dish
through a 60 $\mu$m orifice, forming an ``NUS'' pattern. Using the manually operated stage
and single-drop printing the formation of this entire pattern took about 5 min. As a result,
most of the droplets dried up during the printing, leaving only dead cells inside the dried
outline of the droplets. A viable substrate is necessary in order to maintain suitable
moisture to prevent cell death from dehydration. Figure~\ref{Fig8}b shows the printed cells
inside the crosslinked gel made from 1.0\% (w/v) alginate and 0.5\% (w/v) calcium chloride.
(Alginate was coated onto well-plate before printing, while calcium chloride was mixed
within the cell medium). The overlapping droplets form a continuous straight edged line.
It was subsequently found that fibroblast cells retained their spherical shape rather
than extending filopodia, which meant that the cells failed to adhere to the alginate.
Similar results have been reported before by Kuo and Ma \cite{Kuo2001}.

In Figure~\ref{Fig8}c, the same ``NUS'' pattern was created by dispensing the cells onto a
collagen gel. Printed cells were immediately placed into an incubator. 1 hour following
the printing, fresh medium was added into the well plates. The samples were transferred
into the incubator again and observed under microscope at intervals. After 5 days,
Live/Dead assay was applied to the samples. A bright green fluorescence was observed
after incubation for 30 min. The cells were shown to survive after printing, adhere to
the gel, spread and proliferate, forming a denser pattern. It is worth noting that the
cells were slightly moved from their initial position, perhaps caused by the addition
of the fresh medium, thereby slightly reducing the resolution of the printing.

\section{Discussion}

This study has demonstrated that piezoelectric DOD inkjet printing is able to
successfully deliver L929 rat fibroblast cells through nozzles as small as 36 $\mu$m.
There was no significant cell death when dispensing the cells through the 81 $\mu$m and
the 119 $\mu$m nozzle, with the mean survival rates only reducing from 98\% to 85\%.
This is in good agreement with existing work by Saunders {\it et al.} \cite{Saunders2008} in which a
commercial printer was used to print human fibroblast cells.

When the orifice was reduced to 36 $\mu$m, the corresponding cell survival rates fell
from 95\% to 76\% when the excitation pulse amplitude increased from 60 V to 130 V.
These results indicate that the droplet ejection out of the nozzle has exerted large
shear stresses on the cells and possibly disrupted the cell membrane and killed about
20\% of the cells. The mean shear rate was estimated by combining the effects of droplet
velocity and orifice diameter and was correlated with the cell survival rate.
A wide range of mean shear rates from $1.3 \times 10^4$ $s^{-1}$ to $9.2 \times 10^5$ $s^{-1}$
were produced within the various nozzle sizes and dispensing speeds.
Cell survival rates were found to be strongly affected by the higher
mean shear rates, especially when the shear rate exceeds $5 \times 10^5$ $s^{-1}$.

The distribution of the number of cells within each droplet was also investigated.
This was done to find out the minimal cell concentration in the medium, which is required
to avoid the appearance of empty droplets, since droplets containing no cells may be
detrimental to pattern printing. The distribution of cell numbers is found to have a
binomial form, which consistent with a uniform distribution of cells inside the
medium in the reservoir and lack of coagulation within the printhead.

For pattern printing, L929 fibroblast cells were delivered by using a 60 $\mu$m nozzle.
Printed cells successfully kept their patterns in the crosslinked gel made from
1.0\% (w/v) alginate and 0.5\% (w/v) calcium chloride. However, it was found that the
cells failed to adhere to alginate. On the other hand, cells dispensed onto collagen
gel were found to successfully maintain their viability, adhere to the gel, spread
and proliferate, forming a denser pattern. However, unlike the crosslinked calcium-alginate
which can immobilize cells quite rapidly, cell adhesion to collagen needs a relatively
long time to get established. Therefore, some of the printed cells were slightly moved
from their initial position when the sample was disturbed, by the addition of fresh
medium or unintended shaking of the sample, which will reduce the resolution of the
printing. The smallest nozzle, with orifice diameter of 36 $\mu$m, was not used for
pattern printing, due to issues concerning the reliability of the printing process,
as it can easily get clogged. Future studies will involve experiment with more
mammalian cell types. It is also of interest to check whether adding ligands or
collagen into alginate (before the crosslinking reactions) will promote cell adhesion
onto the substrate.
\clearpage

\begin{figure}[!ht]
\begin{center}
\includegraphics[width=4.5in]{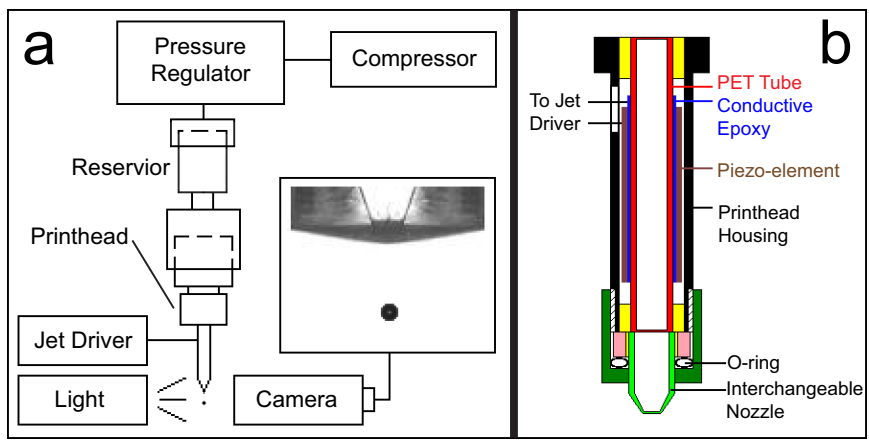}
\end{center}
\caption{
{\bf Schematic showing the drop-on-demand inkjet printing system used in the experiment.}
(a) The overall set-up.
(b) The interchangeable nozzle design \cite{Li2010}.
}
\label{Fig9}
\end{figure}

\section{Materials and Methods}

\subsection{Preparation of Cells, Alginate and Collagen}

L929 rat fibroblasts were cultured in 1 $\times$ Dulbeccos Modified Eagles Medium (D1152).
Medium was supplemented with 10\% foetal bovine serum (FBS, Gibco) and 1\% Penicillin
Streptomycin. Cells were cultured in 150 $cm^3$ culture flasks at 37 $^{\circ}$C, 5\%
CO$_2$ and were observed under a microscope at intervals until they grew to full layer
of the flasks. Cells were then rinsed with phosphate-buffered saline (PBS, Gibco) and
trypsinised using 0.25\%, 1 mM EDTA Na (Gibco). DMEM was mixed with the trypsinised
cell solution and transferred to 50 $cm^3$ conical tubes which were then centrifuged at
1500 rpm for 5 min. After centrifugation, the supernatant liquid was removed leaving
the cell pellet. Fresh media was added and mixed thoroughly using a pipette to obtain
the cell suspension of a specified concentration by using a haemocytometer.

A 1.0\% (w/v) aqueous solution of sodium alginate (A2158, Sigma-Aldrich) was prepared
by suspending the polymer in distilled water. After 6 hours of stirring by a magnetic
stirrer, the solution was sterilized by sterile filtration, using 0.22 $\mu$m membrane
filters.

The 3 mg/ml collagen solution, (C4243, Sigma-Aldrich) was prepared by mixing 8 parts
of chilled collagen solution with 1 part of 10 $\times$ PBS. The pH of mixture was
adjusted to 7.2 to 7.6 using 0.1 M NaOH. The pH value was monitored carefully using
pH paper. To prevent gelation, the resulting solution was then maintained at
temperature of 4 $^{\circ}$C until ready for use.

\begin{figure}[!ht]
\begin{center}
\includegraphics[width=5.0in]{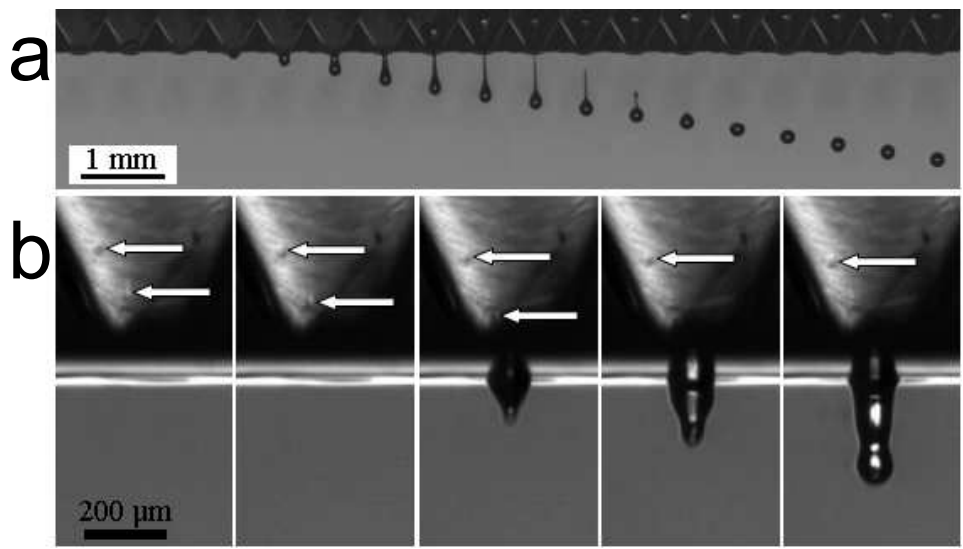}
\end{center}
\caption{
{\bf Droplet ejection images taken by using the high-speed camera.}
(a) Image sequence showing the formation of a 160 $\mu$m droplet from a 119 $\mu$m nozzle,
taken at a frame rate of 8,000 fps, giving time between frames of 125 $\mu$s. Liquid used
was 1.0\% (w/v) aqueous solution of sodium alginate. Drop velocity is 0.74 m/s.
(b) Images showing cell motion in the nozzle. Nozzle diameter was 119 $\mu$m.
}
\label{Fig10}
\end{figure}

\subsection{Printing Setup}

Printing experiments were carried out by using a self-developed squeeze mode piezoelectric
inkjet printing system. The setup is comprised of an air compressor, a pressure regulator, a
reservoir, a piezo-actuated printhead and a piezo driver, as shown in Figure~\ref{Fig9}.
High-speed video images were recorded by a Photron Fastcam SA-1 camera, at a frame rate
of 8,000 fps. Backlighting was accomplished with an Arrisun-5 lamp, which was shone onto a diffuser.
The printhead consists of a glass tube with an annular piezoelectric transducer tightly bonded at its
outer-surface by epoxy. The dispensing nozzle was produced by heating and pulling one end of
another glass tube until it melted and formed a hollow cone with a closed end. The closed end was
polished until an orifice of a desired diameter appeared. The present printhead design is
a great improvement over conventional pintheads, as it allows for the use of interchangeable
nozzles, for the same piezoelectric transducer. The interchangeable nozzle design allows one
to easily clean or change the clogged or damaged nozzle. Figure~\ref{Fig9} also shows this new design
schematically and the details of the design have already been published
elsewhere \cite{Li2010}.

The combination of the compressor and the pressure regulator provides a negative pressure
in the reservoir to hold up and prevent the liquid from leaking out of the orifice of the
printhead. Electric signals are sent by a JetDrive$^{TM}$III (Microfab Technologies Inc.)
to the piezoelectric transducer (Boston Piezo-Optics Inc.), causing alternating expansion and contraction of the
transducer as well as the glass tube, ultimately, squeezing the liquid inside the tube
and ejecting a droplet from the orifice.

The inkjet process is highly periodic. Figure~\ref{Fig10}a shows the droplet formation process in
time sequence. Drop velocity can be calculated by dividing the spacing between droplets in two video frames
by their time difference. Figure~\ref{Fig10}b shows images of a few cells inside the nozzle.

For the study of cell survival rates, L929 rat fibroblast cell suspensions were printed
through orifices of  three different diameters (119 $\mu$m, 81 $\mu$m and 36 $\mu$m) onto
well plates (Costars) which contained the live-dead assay solution. Preparation of the
live-dead assay solution will be introduced later. The electric pulses which were used to
drive the piezoelectric transducer were in the range of 52 to 140 V.  Each sample was
printed for approximately 20 s with a driving frequency of 1.5 kHz for the printhead.
Prior to the printing process, a 15 $\mu$l cell suspension was deposited into a well
plate in the same environment as the printing system, to act as a control.

For pattern printing, either alginate or collagen served as the substrate. The 1.0\% (w/v)
alginate was coated onto well-plate surfaces (Costar) to form around 100-$\mu$m-thick film.
Cells were dispensed onto this film using a cell ink which contained 0.5\% (w/v) calcium
chloride and had a cell concentration about 3 $\times$ $10^6$ cells per ml. The crosslinking
reaction occurs once the droplets contact the alginate film. Printed samples were
immediately placed into an incubator. One hour after printing, fresh medium was carefully
added into the well plates, covering the gel surface and protecting the cells from
dehydration. Samples were transferred into incubator again and observed under a
microscope at intervals. When collagen served as substrate, 0.3\% (w/v) collagen
solution was coated onto well plate surfaces to around 2-mm-thick films and warmed up to
37 $^{\circ}$C for around 1 hour for gel formation. The well plates were then placed onto
a X-Y motion stage and L929 cell suspensions were printed onto the gel according to the
desired pattern. Printed samples were immediately transferred into an incubator. Fresh
medium was carefully added into the well plates 1 hour following the printing, preventing
cells death from dehydration. Samples were transferred into incubator again and observed
under a microscope at intervals.

\subsection{Survivability Tests}

A Live-Dead Viability/Cytotoxicity Kit (L3224, Molecular Probes, Invitrogen) was used
to assess the survivability of the cells after the printing. The frozen vials containing
the assay were thawed and centrifuged briefly before use. 20 $\mu$l of the supplied 2 mM Ethidium homodimer (EthD-1)
solution and 5 $\mu$l of the supplied 4 mM calcein AM solution were added into
10 ml of 1 $\times$ DMEM solution and mixed thoroughly, which gave an approximately 4 $\mu$M
EthD-1 and 2 $\mu$M calcein AM working solution. Cells were directly dispensed into well
plates which each contained 100 $\mu$l of the assay mixture, then incubated for 30 min.
For each printing condition, cells were dispensed into 5 separate petri dishes, to study
the variation of survival rates. Controls were taken directly from the cell ink before
printing and put on a set of separate petri dishes, undergoing the same environment and
procedure. The stained samples were then partly transferred onto microscope slides and
observed under a fluorescence microscope. 6 images were captured from each petri dish
for cell counting. Cells that remained alive after the printing were stained green and
the damaged cells were stained red.  The numbers of alive and dead cells for each sample
were tallied with respect to that of the control which was taken prior to the printing.

\section*{Acknowledgments}

The experimental part of this study was performed while STT and EQL were at National University of Singapore.
The authors would like to thank Prof. Fuh, Jerry Ying Hsi's group for their cooperation
in providing necessary equipments for the experiments. The authors also thank Prof. Evelyn Yim and
Dr. Gajadhar Bhakta, members from the Tissue Engineering lab in National University
of Singapore, for their help in the preparation of the biomaterials. We also appreciate the
support of Prof. Dietmar W. Hutmacher and Benjamin Wu Yong Hao during the early phase of this work.
Funding was provided by A-STAR grant R-265-000-224-305.

\end{document}